\documentclass[aps,reprint,prapplied,superscriptaddress,twocolumn]{revtex4-2}
\usepackage{graphicx}
\usepackage{physics}
\usepackage{amsfonts}
\usepackage{geometry}
\usepackage{siunitx}
\usepackage[T1]{fontenc}
\usepackage{subfigure}
\usepackage{float}
\usepackage[section]{placeins}
\usepackage[hidelinks]{hyperref}

\usepackage{tabularx}
\usepackage{array}
\newcolumntype{C}{>{\centering\arraybackslash}X} % Apparently making a table is such a PITA that we need a custom definition

\usepackage[normalem]{ulem} % If you don't add the [normalem] then it'll ruin your bibliography formatting

\begin{document}
\title{Simultaneous Detection, Demodulation, and Angle-of-Arrival Determination of Communication Signals Using a \\ Dual Ladder Rydberg Receiver}
%\title{A Direct-Conversion Rydberg Receiver for Simultaneous Detection, Demodulation, and Angle-of-Arrival Determination of Communication Signals}

\author{Stone B. Oliver}
\email[]{stone.oliver@nist.gov}
\affiliation{National Institute of Standards and Technology, Boulder, Colorado 80305, USA}
\affiliation{Department of Physics, University of Colorado, Boulder, Colorado 80309, USA}

\author{Samuel Berweger}
\affiliation{National Institute of Standards and Technology, Boulder, Colorado 80305, USA}

\author{Eugeniy E. Mikhailov}
\affiliation{College of William \& Mary, Williamsburg, Virginia 23187, USA}

\author{Dixith Manchaiah}
\affiliation{National Institute of Standards and Technology, Boulder, Colorado 80305, USA}
\affiliation{Department of Physics, University of Colorado, Boulder, Colorado 80309, USA}

\author{Nikunjkumar Prajapati}    	
\affiliation{National Institute of Standards and Technology, Boulder, Colorado 80305, USA}

\author{Christopher L. Holloway}
\affiliation{National Institute of Standards and Technology, Boulder, Colorado 80305, USA}

\author{Matthew T. Simons}
\email[]{matt.simons@nist.gov}
\affiliation{National Institute of Standards and Technology, Boulder, Colorado 80305, USA}

\begin{abstract}
In a typical Rydberg mixer, modulated communication signals are detected using a radio frequency (RF) heterodyne technique. The mixer outputs an intermediate frequency (IF) which must be filtered and mixed down to baseband. In this work, we apply an RF-homodyne technique to simultaneously demonstrate baseband I/Q readout and angle-of-arrival determination of standard communication signals using a dual ladder Rydberg receiver (DLRR). We also provide a systematic comparison with a conventional Rydberg receiver (CRR) (i.e., with a heterodyne-based receiver). While the signal amplitude of the DLRR does not decay with increasing symbol rate as in a CRR, direct conversion receivers are highly sensitive to low-frequency noise. As a result, the DLRR's performance is degraded relative to a CRR when subjected to substantial pink noise. We show that both systems behave comparably once pink noise effects have been accounted for.
\end{abstract}

%In this work, we apply a radio frequency (RF) homodyne measurement technique to demonstrate a direct, baseband readout and simultaneous demodulation of the in-phase (I) and quadrature (Q) components of standard communication signals using a dual ladder Rydberg receiver. We also show that the polarization sensitivity of this receiver design can be used to determine the signal's angle of arrival (AoA). The measured AoA as a function of the signal's true AoA is shown to be a monotonic curve. However, this curve varies sensitively upon the experimental setup's exact geometry. We then compare the dual ladder system with a more conventional RF-heterodyne-based Rydberg receiver. While both systems exhibit comparable performance at lower symbol rates, we find that the dual ladder receiver design is more sensitive to low-frequency noise than its conventional counterpart, which negatively impacts performance at higher symbol rates.
\maketitle

\section{Introduction}
\label{IntroductionSection} 
Rydberg receivers are Rydberg atom-based sensors designed to detect radio frequency (RF) signals \cite{Schlossberger_2024_Gen, Fancher_2021, Shaffer_2019} which offer several key advantages over traditional RF receivers \cite{Meyer_2018, Paul_2018, Song_2019, Holloway_2019, Otto_2021, Liu_2022, Meyer_2023, Yuan_2023, Cai_2023, Shaffer_2024, Gao_2025}. Traditional RF receivers are size-limited due to (1) their use of complicated circuits, amplifiers, mixers, and digitizers, and (2) the fact that the antenna used to receive the signal must have a size comparable to the wavelength of the signal. In contrast, a Rydberg receiver forgoes much of this complicated hardware, and its physical size is not sensitive to the signal's wavelength. Given their reliance on the fixed, discrete energy structure of atoms, Rydberg receivers can also be self-calibrated. Additionally, traditional RF receivers are band-limited, whereas a single Rydberg receiver could plausibly span a frequency range from $\approx \SI{100}{\mega\hertz}$ to $ \SI{1}{\tera\hertz}$.

A Rydberg mixer \cite{Simons_2019}, also known as an atomic super-heterodyne receiver \cite{Jing_2020}, is a phase-sensitive Rydberg receiver capable of detecting standard communication signals (e.g., amplitude phase shift keying (APSK), 16 quadrature amplitude modulation (16QAM), etc.). It is comprised of a two-photon electromagnetically induced transparency (EIT) spectroscopy system and an RF field tuned to a Rydberg-Rydberg atomic transition. In this situation, the RF field produces Autler-Townes (AT) splitting in the EIT feature. Using the RF field as a local oscillator (LO), the Rydberg receiver acts as a free-space mixer which down-converts an incoming RF signal to an intermediate frequency (IF). The IF is then encoded in the EIT signal. Using a spectral filter to remove negative frequency components and mixing the recorded IF down to baseband, the in-phase (I) and quadrature (Q) components of the signal correspond to the real and imaginary parts of the output, respectively. Given its prevalence in the literature, we refer to this receiver design as a conventional Rydberg receiver (CRR).

While a CRR typically relies on the use of RF-heterodyne detection, one can also use RF-homodyne detection, where the LO frequency is matched to the signal frequency. Applying RF-homodyne detection to a CRR, any modulation encoded in the signal is directly converted to baseband, but the output is insensitive to any components of the signal which are out of phase with the LO. For this work, we use 'out of phase' to refer exclusively to two waves being $90^{\circ}$ out of phase with respect to one another. Thus, a CRR can be used in conjunction with RF-homodyne detection to receive amplitude modulated signals which are not out of phase with the LO \cite{Yang_2023}, but it will be generally ineffective in detecting phase modulated signals (i.e., it can measure either I or Q, but not both).

This limitation can be circumvented through the simultaneous use of two independent, spatially overlapped CRRs, where the LO fields of each CRR are out of phase with one another. We refer to this design as a dual ladder Rydberg receiver (DLRR), and we refer to the individual receivers therein as the arms of the DLRR. In this situation, a signal which is out of phase with one arm will be in phase with the other, and vice-versa. As such, one can use RF-homodyne detection with the DLRR to directly measure I with one arm and Q with the other, removing the need to filter out negative frequency components and mix the signal down to baseband. Hence, the DLRR simultaneously detects and demodulates amplitude and phase modulated signals. As with a CRR, the DLRR is also sensitive to the signal's relative phase.

Another advantage of the DLRR is its inherent ability to determine a modulated signal's angle of arrival (AoA) via measurement at a single spatial location. Previous work has demonstrated AoA measurements using a super-heterodyne architecture, but measurements were made using two spatially separated laser beams and measured only a fixed RF tone, not a modulated signal \cite{Robinson_2021, Yan_2025}. Since these results rely on a comparison of phase at two different locations, extension to modulated signals where the phase varies over time requires a nontrivial synchronization of the two measurements. Conversely, the DLRR relies on setting the RF-polarization of each arm's LO orthogonal to the other; the magnitude of the signal recorded by each arm varies naturally based upon the signal's polarization overlap with each LO. The ratio of the signal amplitude recorded by each arm can then be used to determine the signal's AoA.

The maximum detectable symbol rate for each system is also influenced by different factors (the symbol rate is defined as the rate at which the signal is altered by the applied modulation, corresponding to the number of symbols transmitted per second). Within any AT-splitting based Rydberg-atom receiver, the measured signal amplitude is maximal when the signal frequency equals the Rydberg-Rydberg transition frequency in the atoms \cite{Simons_2019, Manchaiah_2026}. Assuming the signal is set to this ideal frequency, the DLRR will then exhibit a maximal amplitude, regardless of the signal's symbol rate. On the other hand, a CRR necessitates the use of an IF, and thus fails to function when either the LO or the signal are detuned such that the amplitude falls below the noise floor of the system. Since the IF must be approximately greater than or equal to the signal's symbol rate, the AT-splitting amplitude contributes to the upper limit on the CRR's maximum detectable symbol rate. Since the DLRR is not subject to this limitation, it is plausible to detect signals at higher modulation rates than achievable with an equivalent heterodyne system (e.g., a CRR). In practice, however, this is difficult to realize since homodyne based systems are inherently sensitive to low-frequency noise \cite{Razavi_2011}.

In this article, we demonstrate the simultaneous detection and direct, baseband readout of the I and Q components of standard communication signals using the DLRR. We also demonstrate the sensor's inherent ability to resolve the signal's relative phase and AoA. Finally, the system's performance is compared with a CRR, and the influence of the type and frequency distribution of noise within each system is discussed. We experimentally demonstrate that the DLRR's performance is degraded relative to the CRR in the presence of multiplicative pink noise. Post-process analysis of these measurements indicates that both systems behave comparably in the absence of multiplicative pink noise.

\begin{figure*}
\includegraphics[width=0.9999\linewidth]{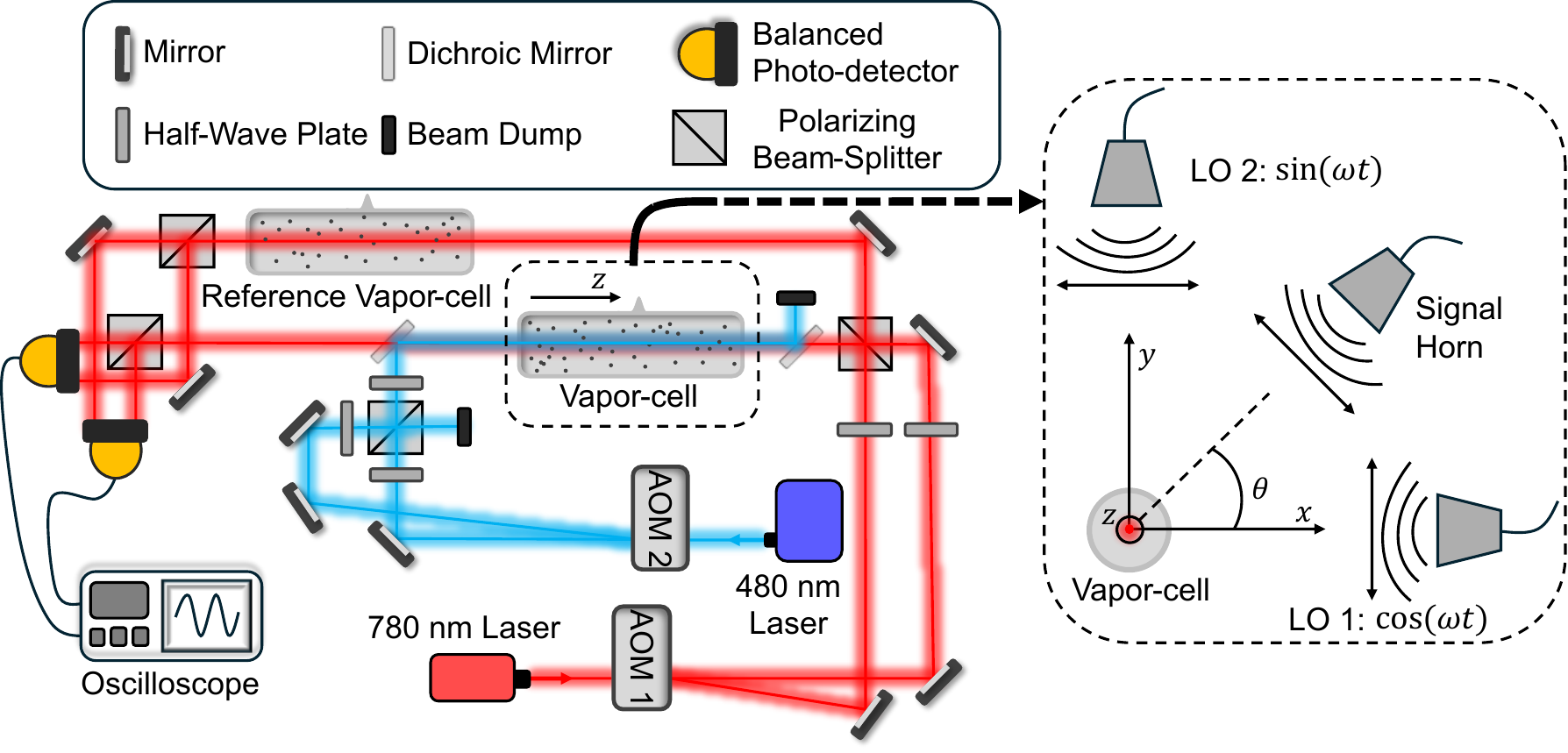}
\caption{Schematic of the optical setup and the orientation of the RF horn-antennas relative to the vapor-cell. The angle $\theta$ describes the relative angle between LO 1 and the signal-horn, i.e., the AoA. The z-direction is taken to be opposite the direction of the probe's propagation at the vapor-cell. Also note that the y-direction is normal to the table on which the experiment was built.}
\label{ExperimentalSetup_Fig}
\end{figure*}

\section{Experimental Setup}
\label{ExperimentalSetupSection}

\subsection{Constructing the \\ Dual Ladder Rydberg Receiver}
\label{System_Construction_Section}

\begin{figure}
\includegraphics[scale=.45]{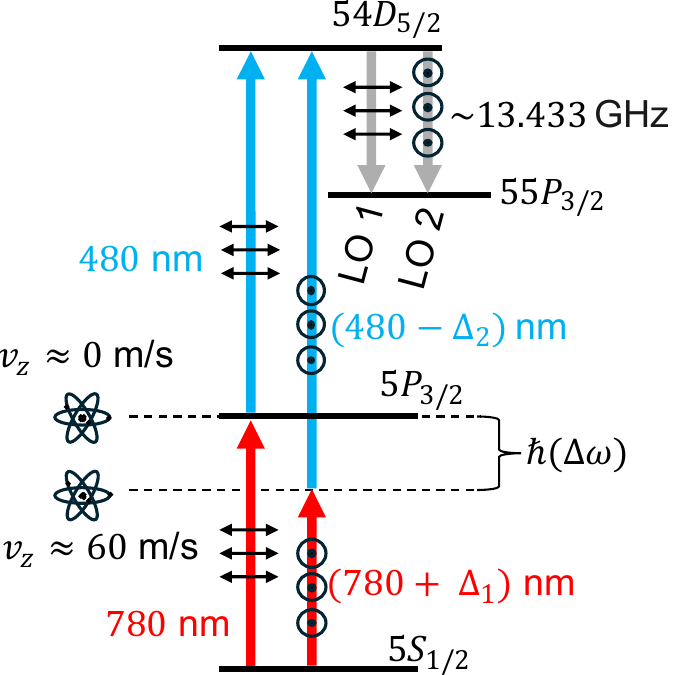}
\caption{Above shows the energy level diagram for $85$-Rubidium used to construct the DLRR. To achieve independent sensing, the second arm interacts with the same states as the first but uses a different Doppler class of atoms and light polarized orthogonally to the first. The atom velocity which shifts the beams to resonance for this work is provided on the left, where the positive z-direction is taken to be opposite the direction of the probe's propagation.}
\label{Energy_Diagram}
\end{figure}

We construct the DLRR similarly to the process outlined in \cite{Berweger_2024}. In order to achieve two spatially overlapped, yet independent CRRs, we begin with two standard, two-photon ladder-type EIT systems \cite{Finkelstein_2023, Derose_2023} which utilize orthogonal optical polarizations and frequencies which are detuned from one another. These choices result in each system interacting with different Doppler classes of atoms, and orthogonal polarizations, thereby eliminating cross-talk between them. The first EIT system is built using a $\SI{780}{\nano\meter}$ probe laser tuned to the $D2$ transition of 85-Rubidium and a $\SI{480}{\nano\meter}$ coupling laser tuned to couple the intermediate $5P_{3/2}$ state to the $54D_{5/2}$ state. Prior to entering a glass vapor-cell containing room-temperature Rubidium atoms, both beams are linearly polarized along the same axis. Additionally, the beams are directed into the cell from opposite directions to partially cancel Doppler broadening effects. EIT is observed by scanning the coupling laser frequency and measuring the probe transmission through the vapor-cell. Figure (\ref{ExperimentalSetup_Fig}) provides a schematic of the experimental setup, while figure (\ref{Energy_Diagram}) depicts the energy level diagram for the EIT setups.

\begin{figure}
\includegraphics[width=0.9999\linewidth]{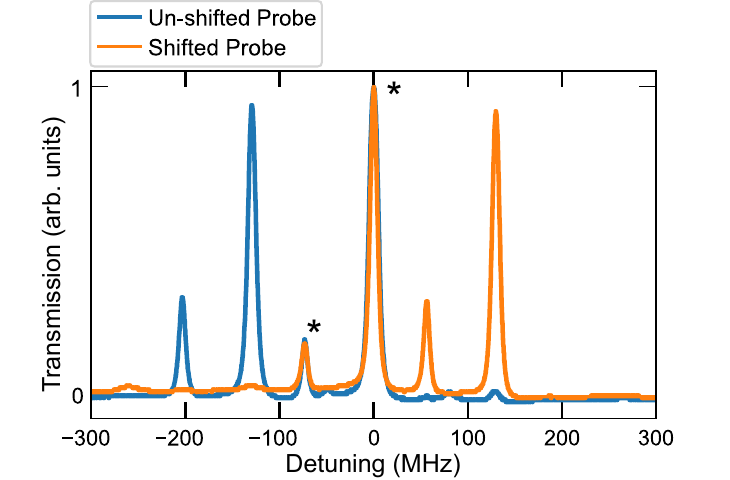}
\caption{Above displays the EIT spectra acquired by measuring the transmission of each probe beam while scanning the coupling beam frequencies. The detuning axis is reported using the un-shifted EIT system. Since all four beams are spatially overlapped, each spectrum shows two pairs of EIT features separated by the detuning between the coupling beams. The smaller peak seen in each pair corresponds to the $54D_{3/2}$ state, while the larger peak corresponds to the $54D_{5/2}$ state. The difference between these peaks was used to calibrate the frequency axis. The features marked with a $*$ correspond to the frequency-overlapped features, the larger of which was used for this work. The unmarked features result from beam cross-talk and do not impact the experiment.}
\label{DualLadder_Spectra}
\end{figure}

In order to construct the second system, the coupling and probe lasers are passed through acousto-optic modulators (AOMs) such that the probe laser is red-shifted by $ - c / \Delta_{1} \approx - \SI{80}{\mega\hertz}$ while the coupling laser is blue-shifted by $+c / \Delta_{2 }\approx + \SI{130}{\mega\hertz}$ where $c$ is the speed of light and $\Delta_{i}$ refers to the wavelength shifts shown in figure \ref{Energy_Diagram} (the original EIT setup is constructed using the $0^{th}$ order light from each AOM). These frequency-shifted beams are then linearly polarized orthogonally to their original counterparts and recombined with those counterparts in order to produce two spatially overlapped EIT setups. While independent sensing along each arm was demonstrated in \cite{Berweger_2024}, this work also tunes the AOMs to ensure frequency-overlapped EIT spectra to allow for simultaneous sensing with both ladders, as seen in figure (\ref{DualLadder_Spectra}), which displays the spectra recorded by each probe laser's transmission. To improve the signal to noise ratio (SNR), we employ balanced detection, where the reference beams for each EIT system are also spatially overlapped and sent through a separate, but identical vapor-cell. For both EIT systems, the probe laser was set to a power of $\approx \SI{100}{\micro\watt}$ with corresponding Rabi rate $\approx \SI{7}{\mega\hertz}$, while the coupling laser was set to a power of $\approx \SI{60}{\milli\watt}$ with a corresponding Rabi rate of $\approx \SI{2}{\mega\hertz}$.
%For both EIT systems, the probe laser is of $1/e^{2}$ radius $\approx \SI{1}{\milli\meter}$ at %the center of the cell and is of power $\approx \SI{107}{\micro\watt}$. The coupling beam's %$1/e^{2}$ radius at the center of the cell is $\approx \SI{.3}{\milli\meter}$ and the power is %$\approx \SI{60}{\milli\watt}$. 

To finish constructing the DLRR, we add to each EIT system an RF horn-antenna tuned to a Rydberg-Rydberg transition in the atoms. The field produced by each horn acts as the LO for its respective arm. For our work, we use the $54D_{5/2}$ to $55P_{3/2}$ transition, which was calculated to be $\SI{13.433374}{\giga\hertz}$ using the Alkali-Rydberg calculator in Python \cite{ARC_2017}. Three conditions must be met when placing these horns. First, each horn must be placed such that the atoms are in the Fraunhofer region of the radiation field (i.e., the radiative far field region), meaning the radiation may be treated as a plane-wave with a well-defined polarization. Second, both horns must be oriented such that their polarizations match the polarization of their respective EIT beams. Third, the field produced by one horn-antenna at the vapor-cell must be out of phase with the field produced by the other horn (i.e., the LO fields must be out of phase with one another).

\begin{figure*}
\includegraphics[width=0.9\linewidth]{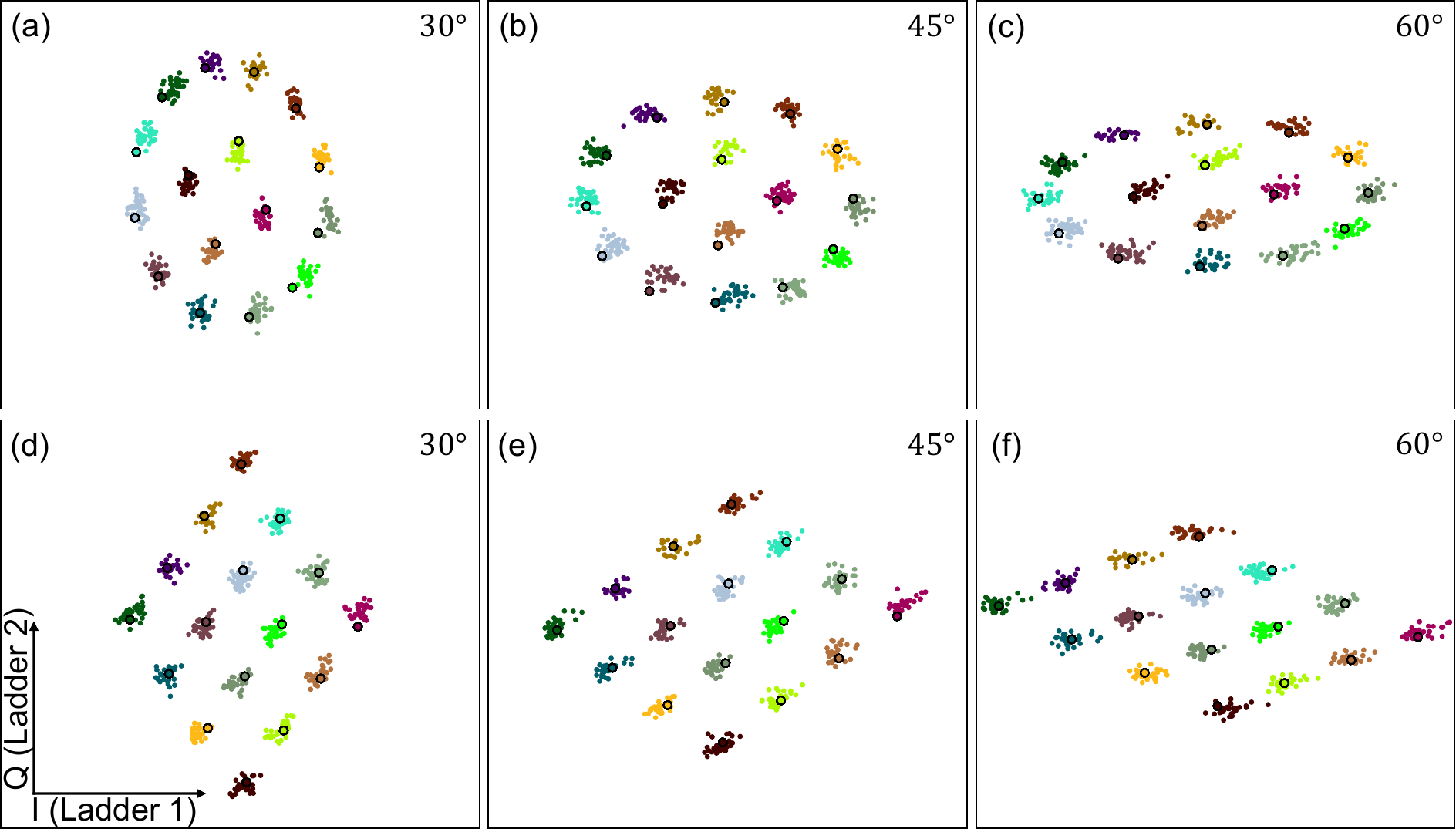}
\caption{Above shows three APSK and three 16QAM I/Q diagrams as a function of the received signal's measured AoA; each constellation is compressed or expanded along each axis according to equation \ref{IQ-WarpEqn}. The individual points in each sub-plot are measured data, while each black circle with a distinct color inside represents a nominal constellation point. (a) APSK with AoA $=30^{\circ}$. (b) APSK with AoA $=45^{\circ}$. (c) APSK with AoA $=60^{\circ}$. (d) 16QAM with AoA $=30^{\circ}$. (e) 16QAM with AoA $=45^{\circ}$. (f) 16QAM with AoA $=60^{\circ}$.}
\label{ConstellationPlotsImage}
\end{figure*}

\subsection{Generating and Detecting a Signal}
\label{Generating_Modulated_Signals_Section}

To complete the experimental setup, we add a third RF horn-antenna in order to provide the signal which the DLRR will detect and demodulate. As with the LO horn-antennas, it must be placed sufficiently far from the vapor-cell. To generate a modulated signal, we use an arbitrary waveform generator (AWG) set to output a modulated tone. Using either a 16QAM, APSK, or quadrature phase shift keying (QPSK) modulation scheme, we select the desired symbol rate and generate a pseudo-random (PN) sequence of 511 symbols (i.e., we generate 511 constellation points, where the duration of the transmitted signal is given by 511 divided by the symbol rate). Since the AWG used for this work cannot directly produce modulated signals at $\SI{13}{\giga\hertz}$, we mix the AWG output with a pure RF-tone using a conventional mixer to produce a modulated signal at the desired frequency. The frequency of the resulting signal is given by
\begin{equation} \label{signalFrequencyEqn}
f_{c} + f_{p} = f_{s},
\end{equation}
where $f_{c}$ refers to the modulated carrier frequency produced by the AWG, $f_{p}$ refers to the pure RF-tone, and $f_{s}$ is the modulated signal frequency. For this work, we set $f_{c} = \SI{1}{\giga\hertz}$ and $f_{p} = \SI{12.433374}{\giga\hertz}$, yielding a modulated signal at frequency $f_{s} = \SI{13.433374}{\giga\hertz}$. The signal is finally sent to and emitted from the signal-horn antenna onto the atoms. The signal is then encoded in the transmission profile of the probe beams, which first pass through the atoms, and then are sent into photo-detectors. The output voltage of these detectors is measured using an oscilloscope.

When the signal field is incident at angle $\theta = 45^{\circ}$ relative to each LO field, it will interact equally with both arms because it possesses equal polarization components along each axis (see figure (\ref{ExperimentalSetup_Fig})). As a result, the relative amplitudes of the measured I and Q values will match their nominal values. As $\theta$ deviates from $45^{\circ}$, the polarization overlap of the signal with each arm will vary, resulting in a shift in I and Q given by
\begin{equation} \label{IQ-WarpEqn}
\begin{split}
I(\theta) &= I_{0}sin(\theta) \\
Q(\theta) &= Q_{0}cos(\theta),
\end{split}
\end{equation}
where $I_{0}$ and $Q_{0}$ refer to the nominal values of I and Q for the given transmission (i.e., the values of $I$ and $Q$ expected when $\theta = 45^{\circ}$). To explicitly depict the way in which the angle of incidence warps the constellation diagram, figure (\ref{ConstellationPlotsImage}) shows three APSK constellation plots and three 16QAM constellation plots collected using signal-horn angles $\theta = 30^{\circ}, 45^{\circ},$ and $60^{\circ}$. Comparing the ratio $I_{0}/Q_{0}$ with the ratio between the relative signal amplitudes measured along each arm of the receiver, $I(\theta)/Q(\theta)$, one can deduce the signal's AoA. Note that equation \ref{IQ-WarpEqn} is idealized and does not consider effects such as RF-reflections or imperfect polarization.

\subsection{Analyzing the Detected Signal}
\label{SignalAnalysisSection}

In order to analyze a given measurement, we first apply a low-pass spectral brickwall filter to the signals recorded by the oscilloscope to remove all frequency components with a frequency larger than the symbol rate used for that measurement. Note, however, that the DLRR does not strictly require the use of a filter to determine I/Q values, as there is no need to remove negative frequency components as is the case with RF-heterodyne detection. To determine the I/Q position of the n$^{th}$ symbol, we then choose to take the value measured at the center point of the symbol's duration.

Once the data is filtered and analyzed, we determine the received signal's relative phase and AoA by comparing to a reference list of nominal I and Q positions for the given PN-sequence and modulation scheme as a function of the signal's phase and AoA. Each list is comprised of $90 \times 360=32400$ entries, corresponding to calculations of the nominal I/Q positions in one-degree increments over the unit circle for phase and over the range $0^{\circ}$ to $90^{\circ}$ for AoA. Subtracting each nominal point set from the measured points to produce a set of error functions, the phase and AoA of the nominal point set which produces the error function of minimal magnitude is taken to be the measured signal's phase and AoA. Using the nominal point set, we also calculate error-vector-magnitude (EVM), where the root-mean-square magnitude of the nominal points is used for normalization \cite{Shafik_2006}.

 For all data related to the signal's AoA and/or phase presented in this work, we apply a symbol rate of 10 kbaud. For data related to system comparison with a CRR, the signal-horn was placed at an angle of $\theta = 45^{\circ}$ relative to both LOs. Lastly, note that when using the DLRR to detect these signals, we tune the power applied to both the LOs and the signal in order to optimize the SNR. For this work, each LO was set to an output power of $\approx$ -16 dBm, while the signal power was $\approx$ -6 dBm. These powers were fixed as the symbol rate was varied.

 Notably, the use of a matched filter is often desirable because it maximizes SNR \cite{Richards_2005}. A matched filter allows one to extract I and Q via maximization of the cross-correlation between the expected signal and the received signal. Furthermore, the use of a spectral brickwall filter on a modulated signal comprised of square pulses inherently leads to error because the filter truncates a portion of the signal's frequency distribution. However, this choice is motivated by the fact that a matched filter is an idealized filter for systems that experience primarily additive noise. As will be discussed in the next section, the DLRR is subject to substantial multiplicative noise (i.e., noise which scales with the signal amplitude). The presence of multiplicative noise renders a matched filter less effective; correlating a received, noisy signal with the nominal signal cannot remove noise which also correlates with the nominal signal. As the use of a spectral brickwall filter is equivalent to convolving the received signal with a sinc function, cross-correlation no longer serves as an effective mechanism for determining I and Q. By instead choosing to only select the center point of the symbol's duration, error induced by the brickwall filter is reduced to a scaling error which is corrected by normalizing the data to the mean of its absolute value. Indeed, the accuracy of this approach was found to be substantially better than the accuracy obtained through the use of a matched filter. One other advantage of this analysis method is that the use of a spectral brickwall filter only requires knowledge of the symbol rate, which can be effectively estimated without any prior knowledge of the signal \cite{Chan_1997, Ciblat_2002, Elgenedy_2013, Majhi_2014}.

\section{Results and Discussion}
\label{ResultsSection}

\subsection{Determining Angle of Arrival and Relative Phase}
\label{AoA_ResultsSections}

The AoA data presented in this section is the result of five separate measurement runs. Two measurements used a 16QAM modulation, two used APSK, and one used QPSK. For this work, the use of 16QAM and APSK modulation schemes provides supporting evidence that our system operates in a linear regime. The ideal choice of modulation scheme is dictated by both system constraints and the operating environment. For example, APSK is favorable when working with low-power systems which rely on non-linear amplification \cite{DeGaudenzi_2006}, whereas 16QAM maximizes the Euclidean distance between constellation points, making it more appealing for linear systems because it requires a lower resolution to function. It is also more robust to fading effects \cite{Zhao_2023}. Looking to figure \ref{ConstellationPlotsImage}, the fact that the APSK and 16QAM measurement results are qualitatively identical indicates that the receiver is operating in a linear regime. In this situation, each modulation scheme may be used interchangeably without any shift in the system's fundamental behavior. We also use QPSK for its lower EVM requirement as it allows for the use of higher symbol rates (see section \ref{ComparisonSection}).

\begin{figure}
\includegraphics[width=0.9999\linewidth]{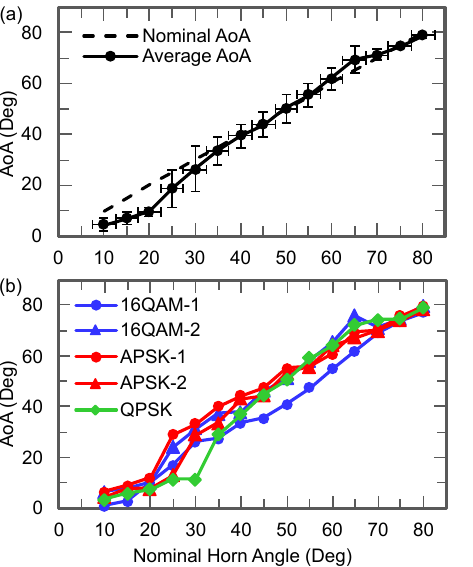}
\caption{(a) displays the AoA found by averaging all five measurement runs. Vertical error bars represent one standard deviation, while the horizontal error bars represent absolute limits on the signal-horn position. (b) displays the phase-averaged AoA measurement results of each measurement run. The color coding used here to represent different modulation schemes is maintained throughout this work.}
\label{AoA_Data_plot}
\end{figure}

For each measurement run, we vary the signal-horn's position from $\theta = 10^{\circ}$ to $\theta = 80^{\circ}$ in $5^{\circ}$ increments. At each angle, we vary the signal's relative phase in 30-degree increments around the unit circle, yielding twelve measurements at a given AoA, where each individual measurement therein is an average of 16 traces from an oscilloscope. The oscilloscope was set to collect 12.5 MSa/s. Averaging over the five measurement runs, the EVM was found to be approximately $15 \pm 3\%$. Figure (\ref{AoA_Data_plot}a) shows the average of all five measurement runs while figure (\ref{AoA_Data_plot}b) shows the phase-averaged AoA versus the signal horn's true placement for each measurement run. 

The average measured AoA well-matches the signal-horn's angle for angles above approximately $25^{\circ}$ (i.e., $25^{\circ}$ above the horizontal plane of the optics table). The deviation at lower horn angles is at least partially a result of polarization-based cross-talk. While RF-absorbing material was used to insulate the vapor-cell from the environment, reflections from the glass walls of the cell itself are well-known to produce standing-wave fields when that cell has physical size comparable to the RF wavelength \cite{Fan_2015}. Indeed, for this work, the cylindrical vapor-cell's diameter was $\approx \SI{2.5}{\centi\meter}$ and the RF wavelength was $\approx \SI{2.2}{\centi\meter}$. Such fields vary appreciably in amplitude over the width of the cell and are well-known to affect AoA measurements \cite{Robinson_2021, Yan_2025}. In the case of the DLRR, these fields induce position-sensitive offsets in both I and Q (stray fields with y-polarization affect I while x-polarized fields impact Q). The resulting error in the I/Q ratio in turn affects the measured AoA. To demonstrate this, the vapor-cell was translated from its optimal position along the y-direction (see figure \ref{ExperimentalSetup_Fig}) by $\pm \SI{2}{\milli\meter}$, and was found to yield a systematic shift in the measured AoA of approximately $\pm 8^{\circ}$ (other attributes such as EVM were unaffected).

This cross-talk error was empirically minimized by optimizing a 16QAM signal's observed constellation diagram at $\theta = 45^{\circ}$ through adjustment of the cell's x-y position. This optimized position is a function of the exact vapor-cell geometry and the RF wavelength, and does not necessarily correspond to the central axis of the cell. Without axial symmetry, changes in the signal horn's angle $\theta$ do not strictly correspond to equal changes in the amplitude of the stray fields induced by the signal along the x and y directions. Thus, changes in $\theta$ are not expected to strictly correlate with equation \ref{IQ-WarpEqn}.

While stray fields resulting from cell-wall reflections cannot be removed, it was also shown in \cite{Fan_2015} that the use of a vapor-cell which is small compared to the RF wavelength induces a more homogeneous stray-field throughout the cell. While not implemented in this work, the use of a smaller cell would reduce the observed sensitivity to cell position because the offsets in I and Q would be rendered fixed. Hence, the resulting system would be more robust against geometric perturbations (e.g., vibrations). Also note that regardless of sensitivity to cell position, the existence of stray fields, homogeneous or not, necessitates the use of a reference curve for calibration when performing measurements in an uncontrolled environment (e.g., use of the data in figure (\ref{AoA_Data_plot}a)).

\begin{figure}
\includegraphics[width=0.9999\linewidth]{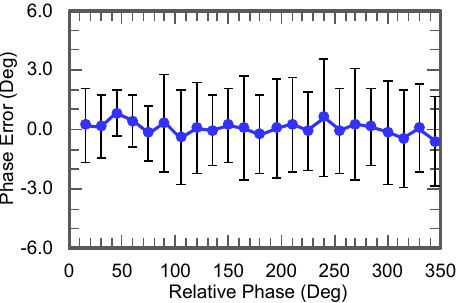}
\caption{Above shows the deviation of the measured relative phase of a 16QAM signal from its nominal value, averaged over signal-horn angles ranging from 10$^{\circ}$-80$^{\circ}$ in 5$^{\circ}$ increments. The error bars represent one standard deviation.}
\label{PhaseData_AoAAveraged}
\end{figure}

We also investigate the DLRR's ability to determine a signal's relative phase. We apply a 16QAM modulation and take the phase of the first measurement to be zero. We then vary the phase in 15$^{\circ}$ increments over the unit circle, yielding twenty-four measurements at a given horn angle. As with the AoA data discussed above, we vary the horn-angle over a range of $\theta = 10^{\circ}$ to $\theta = 80^{\circ}$, evenly spaced by 5$^{\circ}$ increments, yielding a total of 344 relative phase measurements (the measurement taken at horn angle $10^{\circ}$ and relative phase $120^{\circ}$ was discarded due to a procedural error involving the phase of the applied RF). We then determine the AoA-averaged relative phase error, depicted in figure (\ref{PhaseData_AoAAveraged}). The average relative phase error was found to be approximately $0^{\circ} \pm 2^{\circ}$.

\subsection{Comparison of the Direct Conversion Rydberg Receiver with a Conventional Rydberg Receiver}
\label{ComparisonSection}

We also compare the DLRR with a CRR. To produce a CRR with comparable characteristics, we  turn off LO 2 (see figure (\ref{ExperimentalSetup_Fig})) and detune the signal frequency $f_{s}$ in equation \ref{signalFrequencyEqn} from the remaining LO to generate an IF (i.e., a beatnote). This is achieved by detuning $f_{p}$. No other characteristics of the system are altered. Since the IF must be approximately equal to or larger than the symbol rate, we set the IF equal to the symbol rate. We then apply a QPSK modulation, and the modulated IF is detected by measuring the un-shifted probe transmission. We also mix the signal down to baseband prior to applying a low-pass spectral brickwall filter.

\begin{figure}
\includegraphics[width=0.9999\linewidth]{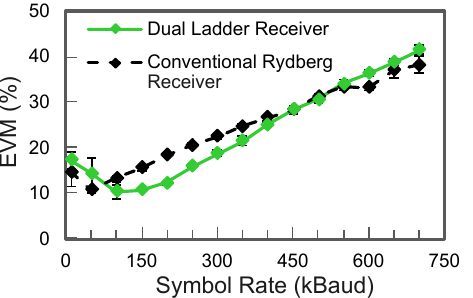}
\caption{Above plots estimated signal EVM versus symbol rate for both the DLRR (homodyne) and the CRR (heterodyne), where multiplicative noise effects have been removed. Both measurements used QPSK modulation. For the CRR, the IF was set equal to the symbol rate. Each data point is an average of five measurements. The error bars represent one standard deviation.}
\label{ComparisonFigure}
\end{figure}

To compare the DLRR with the CRR, we vary the modulation's symbol rate and determine the EVM of the recorded signal. The photo-detectors used for this work possessed a bandwidth of $\SI{750}{\kilo\hertz}$, placing an upper limit on the maximum detectable symbol rate. As the intensity spectrum of a laser is well-known to exhibit $1/f$ noise \cite{Kitching_1993}, we expect the DLRR to exhibit higher EVM owing to the signal's frequency distribution being centered at $\SI{0}{\hertz}$. Since the measured signal amplitude scales directly with probe intensity, this pink noise is also multiplicative in nature. Indeed, the noise spectrum was found to be dominated by multiplicative pink noise, and the DLRR was found to exhibit a larger EVM than the CRR for symbol rates higher than approximately $\SI{300}{\kilo\hertz}$ (see appendix \ref{Noise_Appendix}). The exact frequency where this deviation begins depends upon the spectral overlap between noise and signal. This is determined by (1) the width of the pink noise profile, and (2) where the signal's frequency spectrum is centered. The CRR reduces this overlap by centering the signal at the IF rather than $\SI{0}{\hertz}$. However, we expect both systems to behave comparably once pink noise is accounted for. This behavior is displayed in figure (\ref{ComparisonFigure}), which plots an estimate of EVM versus symbol rate for both systems after the removal of multiplicative noise. Both the total, experimentally measured EVM (i.e., EVM where multiplicative effects have not been removed) and the process of estimating and removing multiplicative noise effects are discussed in appendix \ref{Noise_Appendix}.

\section{Conclusion}
\label{Conclusion}

\begin{table*}
\includegraphics[width=0.7\linewidth]{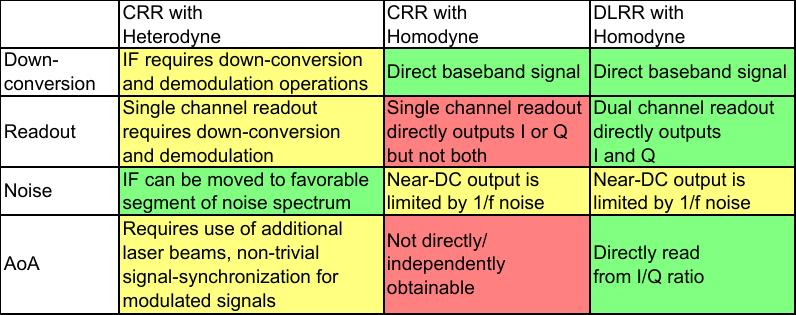}
\caption{The table above summarizes the advantages and disadvantages of the CRR and DLRR architectures. Green cells denote advantageous characteristics, yellow cells denote neutral or potentially disadvantageous characteristics, and red cells denote disadvantageous characteristics. The DLRR architecture is well suited to environments in which $1/f$ noise can be effectively mitigated, although, most practical operating environments exhibit significant $1/f$ noise. While a DLRR employing heterodyne detection could, in principle, estimate a signal's AoA in the same manner as its homodyne counterpart, it offers no other advantages over a CRR employing heterodyne detection. Consequently, it has been omitted.}
\label{Ref_Table} 
\end{table*}

In this work, we have demonstrated a direct, baseband detection and simultaneous demodulation of the I and Q components of 16QAM, APSK, and QPSK signals through the use of a DLRR which utilizes RF-homodyne detection. In addition to determining a signal's relative phase, we have also shown that this sensor design inherently allows for determination of a modulated signal's AoA via measurement at a single spatial location. On the other hand, previous AoA demonstrations in CRRs relied on comparing phase at two locations, a method that does not trivially extend from pure RF tones to modulated signals. While the averaged measurement results displayed a monotonic relationship between measured versus nominal AoA, the measured AoA deviates substantially from the nominal value below approximately 25$^{\circ}$. This deviation is thought to result from RF-reflections inside the glass vapor-cell used for the experiment which cause cross-talk between each arm of the DLRR, meaning the measurement is sensitive to the exact geometric configuration of the setup. However, this issue could plausibly be mitigated through the use of a vapor-cell which is small compared to the RF wavelength. The result is a more homogeneous stray field inside the cell; while this extraneous field cannot be removed, its homogeneity would yield a system which is more robust against geometric perturbations.

Additionally, the performance of the DLRR was compared to a CRR through the use of a QPSK signal with varying symbol rates. Measurements made using the CRR used an IF set equal to the modulation's symbol rate. After the removal of multiplicative pink noise effects (see appendix \ref{Noise_Appendix}), the DLRR showed comparable estimated EVM to the CRR. However, the total, experimentally measured EVM of the DLRR is larger than the EVM observed in the CRR at higher symbol rates. This indicates that sensor performance is influenced by both the type (additive versus multiplicative) and frequency distribution (pink versus white) of noise present in each system. Table \ref{Ref_Table} offers a general comparison of the CRR and DLRR systems. The substantial quantity of multiplicative pink noise in these systems also justified the use of a low-pass spectral brickwall filter followed by the extraction of I and Q via use of the center point of each symbol's duration. It remains theoretically plausible that a DLRR with minimized noise could outperform its conventional counterpart, owing to the fact that a DLRR does not require the use of an IF, meaning the maximum detectable symbol rate is not limited by the AT-splitting amplitude, which begins to decay when the detuning of incident RF-radiation is on the order of $\SI{1}{\mega\hertz}$ \cite{Manchaiah_2026}. On the other hand, this result would be difficult to realize owing to the well-known noise sensitivity of homodyne detection.

Future work may include a systematic investigation and reduction of experimental noise to study the interplay between the functionality of each sensor design and noise present in each system or its environment. A substantial reduction in noise could plausibly enable (1) the use of modulation schemes which require lower EVM than the values reported in this work (e.g., 64QAM), (2) a field test of this receiver design in a more realistic environment, and/or (3) a demonstration of the reception of signals with higher symbol-rate than currently achievable in any existing Rydberg receiver.

\section*{Acknowledgments}
The authors would like to thank Rajavardhan Talashila for valuable simulation results, as well as Daniel Hammerland and Link Patrick for advice and support related to the optical setup. The authors would also like to thank Joshua Kast for valuable conversation related to I/Q demodulation.

This research was developed with funding from NIST. The views and opinions of authors expressed herein do not necessarily state or reflect those of the United States Government or any agency thereof. A contribution of the U.S. government, this work is not subject to copyright in the United States.

\section*{Conflict of Interest}
The authors have no conflicts to disclose.

\section*{Data Availability Statement}
All of the data presented in this paper is available at https://doi.org/10.18434/mds2-4053

\appendix
\section{System Diagnostics, Calibration, and Geometric Sensitivity of the DLRR}
\label{Diagnostics_Section}
\begin{figure}
\includegraphics[width=0.9999\linewidth]{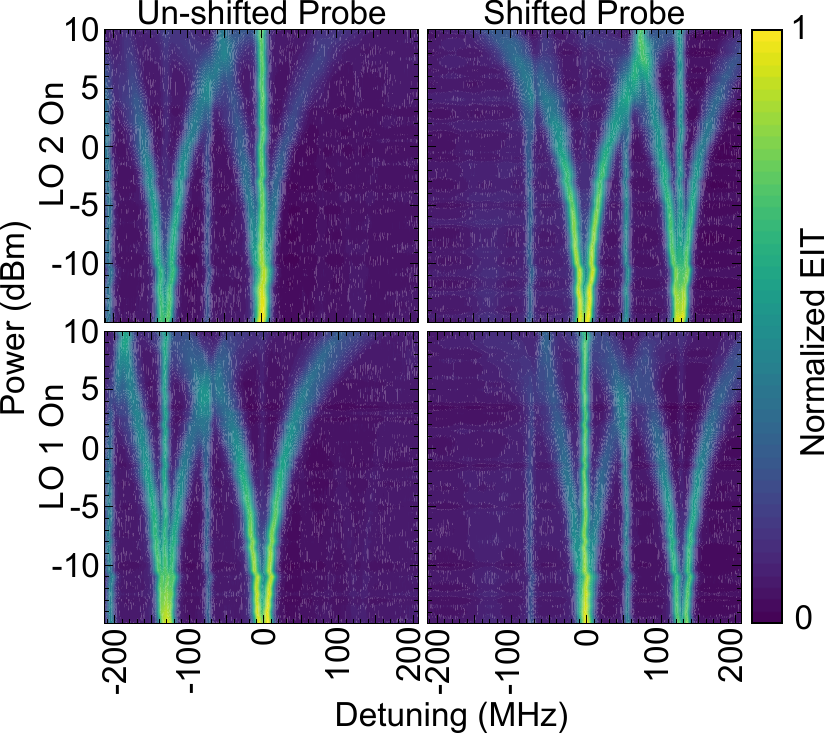}
\caption{Above shows waterfall plots of the probes' transmission amplitudes as a function of both LO power and coupling laser detuning from the center of the EIT feature. Clearly, one arm's EIT feature splits in the presence of an applied field from its corresponding LO, while the other arm's EIT feature still exhibits a distinct peak at zero detuning. The roles reverse upon switching LOs.}
\label{RF_PowerSweep_Plots}
\end{figure}

To verify independent sensing along each arm of the DLRR, we measure each arm's response to each LO, separately. Following the procedure outlined in reference \cite{Berweger_2024}, we turn LO 1 on while LO 2 and the signal-horn are both left off and measure the resulting EIT spectra. We then repeat this process, except we turn on LO 2 while LO 1 is left off. Since the polarization of each LO and its corresponding EIT system are matched, AT-splitting of the EIT feature occurs in one arm, but not the other. Swapping which LO is on, the AT-splitting moves from one arm to the other. Fitting two Gaussian curves to the split peaks, one can deduce the magnitude of the electric field experienced by the atoms; the field strength scales linearly with the square root of the power of the LO horn-antenna. This diagnostic offers a straightforward method of measuring the magnitude of unwanted cross-talk between each arm of the receiver. Figure (\ref{RF_PowerSweep_Plots}) shows the probe beams' transmission spectra as a function of which LO is used and the power of that LO. Given the sensor must operate with the coupling lasers fixed at zero detuning from the overlapped EIT peaks in figure (\ref{DualLadder_Spectra}), these plots demonstrate independent sensing along each ladder, with sufficiently small cross-talk between them when one LO is left off. However, as discussed in the main text, the system's performance was found to be dependent upon the exact position of the vapor-cell, an effect which could be mitigated by the use of a vapor-cell which is small compared to the RF wavelength. 

% Leaving this out as I can't find a good ref on AOM distortion

%In addition to the positioning issue discussed above, it is %also noteworthy that the use of single-pass AOMs is well-%known to distort beam profiles. As such, the regions of %highest intensity for each arm may not perfectly coincide, %potentially contributing to cross-talk in the system as %each region may experience different RF-fields. This could %be remedied through the use of polarization maintaining %fibers placed after the AOMs, which would yield Gaussian %modes for all beams. Although, this may not be feasible if %the power loss resulting from the use of fibers would %reduce the coupling beams' powers to below their desired values.

\section{Multiplicative Noise in Rydberg Receivers}
\label{Noise_Appendix}

\begin{figure}
\includegraphics[width=0.9999\linewidth]{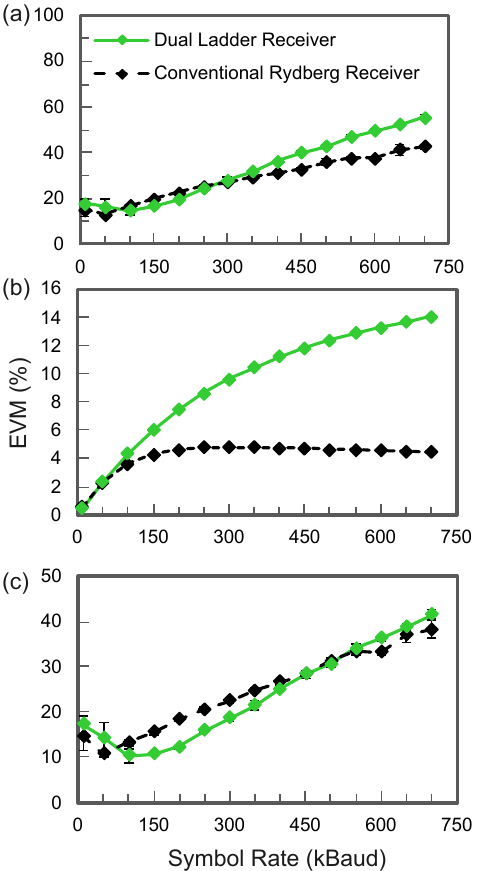}
\caption{(a) plots signal experimentally measured EVM versus symbol rate for both the DLRR (homodyne) and the CRR (heterodyne), where no noise components have been removed. Both measurements used QPSK modulation. For the CRR, the IF was set equal to the symbol rate. Each data point is an average of five measurements. The error bars represent one standard deviation. (b) plots the estimated EVM which results from multiplicative noise effects, such as the probe laser's $1/f$ noise. (c) plots the estimated residual EVM which results primarily from additive noise. This plot is identical to figure (\ref{ComparisonFigure}).}
\label{ComparisonFigure_Full}
\end{figure}

\begin{figure}
\includegraphics[width=0.9999\linewidth]{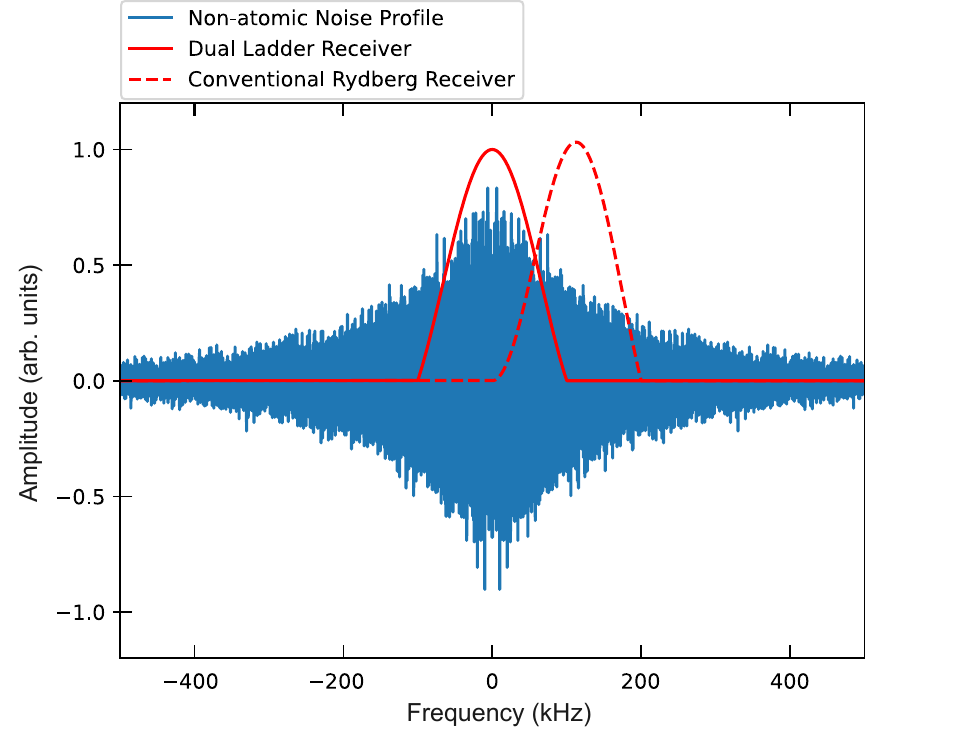}
\caption{Above plots the recorded non-atomic noise profile detected by the CRR (blue) and the envelope of the nominal frequency distribution of the received signal expected using either the CRR (dashed red) and the DLRR (solid red) after truncation with a spectral brickwall filter such that only the primary lobe of the envelope remains (compare with equations \ref{DLRR_EnvelopeEqn} and \ref{CRR_EnvelopeEqn}). The noise profile seen for the DLRR is functionally identical. The nominal frequency distributions were constructed using a symbol rate $S = \SI{100}{\kilo\hertz}$.}
\label{NoiseProfile_WithSincs_Figure}
\end{figure}

Figure (\ref{ComparisonFigure}) displays comparable EVM between the DLRR and CRR only after the removal of multiplicative noise effects. Figure (\ref{ComparisonFigure_Full}a) instead shows the experimentally measured EVM, without the removal of any noise contributions. The clear difference between these two plots highlights the effect of the multiplicative pink noise noted in section \ref{ComparisonSection}. In this section, we address this discrepancy directly through an analysis of how the frequency distribution and type of noise each contribute to EVM. Figure (\ref{ComparisonFigure_Full}b) plots an estimate of EVM which results from multiplicative noise, and figure (\ref{ComparisonFigure_Full}c) is a reproduction of figure (\ref{ComparisonFigure}) for direct comparison. Figure (\ref{NoiseProfile_WithSincs_Figure}) displays the frequency distribution of noise measured by the CRR with no signal present. The noise profile recorded for the DLRR is functionally identical. 

In addition to this noise, the atom response time and the amplitude of the AT-split feature also contribute to EVM, as the AT-splitting amplitude affects the SNR, and the atom response time can affect the system's ability to transition between symbols \cite{Shaffer_2022}. We refer to these additional effects collectively as atomic noise, as these effects fundamentally result from light-matter interaction inside the cell. We break the EVM into two components; one results from the non-atomic noise profile, the other results from atomic noise. We express this as a function of the symbol rate:
\begin{equation}
\text{EVM} = E_{0}(S) + E(S),
\end{equation}
where $E_{0}(S)$ refers to atomic noise induced EVM, and $E(S)$ refers to non-atomic noise induced EVM. Since both systems utilized the same physical setup, differing only by the use of a second LO and an IF, we take $E_{0}(S)$ to be identical for both systems. Thus, the difference in EVM seen in figure (\ref{ComparisonFigure}) is entirely due to $E(S)$. We further assume that the non-atomic noise is primarily multiplicative, as residual $1/f$ laser intensity fluctuations are well known to affect the performance of Rydberg sensors \cite{Tang_2025}. In this situation, $E(S)$ results from complex valued noise described by
\begin{equation}
N(t) = N_{M}(t) + N_{A}(t) \approx N_{M}(t),
\end{equation}
where the subscripts $M$ and $A$ are used to denote multiplicative and additive noise effects, respectively, and the real and imaginary parts of $N(t)$ refer to noise in $I$ and $Q$, respectively. Since $N_{M}(t)$ scales with the amplitude of the received signal, $E(S)$ is expected to scale with the overlap of the non-atomic noise profile with the nominal signal distribution. The envelope of the frequency distribution of both the real and imaginary components of the nominal signal expected from the DLRR is
\begin{equation} \label{DLRR_EnvelopeEqn}
FT[A(t,S) ] = \tilde{A}(f, S) = \frac{S}{\pi f} sin(\pi f/S),
\end{equation}
where $FT[A(t, S)]$ is the Fourier transform of the nominal signal envelope, $A(t, S)$, with respect to time $t$. We also use $\omega = 2 \pi f$ to express functions in the frequency domain using units of Hz. Prior to mixing down to baseband, the envelope of the frequency distribution of the signal expected from the CRR is
\begin{equation} \label{CRR_EnvelopeEqn}
\begin{split}
\tilde{A}(f, S) =& \frac{S}{\pi(f-S)} sin(\pi(f-S)/S) \\
& \pm \frac{S}{\pi(f+S)} sin(\pi(f+S)/S),
\end{split}
\end{equation}
where the second term is positive for the real part and negative for the imaginary part. The real parts of both envelopes are plotted in figure (\ref{NoiseProfile_WithSincs_Figure}) against the noise profile with $S= \SI{100}{\kilo\hertz}$ after truncation with the appropriate spectral brickwall filter. Using the convolution theorem, we cross-correlate the quantity $A(t,S)$ with the error $N(t)$ via multiplication in the Fourier domain:
\begin{equation}
\label{Multiplicative_EVM_EQN}
\begin{split}
FT[A(t, S)\star N(t)] & = FT[A^{*}(-t, S) * N(t)] \\
& = \tilde{A}^{*}(f, S) \tilde{N}(f), \\
\end{split}
\end{equation}
where $\star$ denotes cross-correlation, $*$ denotes convolution, and $\tilde{N}(f)$ is the Fourier transform of $N(t)$. The absolute value of this quantity (i.e., $| \tilde{A}^{*}(f, S) \tilde{N}(f)|$) allows one to quantify the relative magnitude of multiplicative noise between measurements at a given frequency.

While this work calculated EVM by taking only the center point of each symbol's duration, one might more generally wish to average over some window of the symbol's duration centered on the symbol's midpoint \cite{Sadr_1987}. To describe how multiplicative noise contributes to EVM, we more generally consider how averaging over this window alters the calculated EVM. The average noise over a window of the n$^{th}$ symbol's duration centered on the symbol's temporal midpoint is described by
\begin{equation}
\begin{split}
N_{n} & = \frac{1}{T-\Delta} \int_{t_{n} - (T-\Delta)/2}^{t_{n}+(T-\Delta)/2} N(t)dt  \\
& = \frac{1}{T-\Delta} \int_{- \infty}^{\infty} N(t)(H(t-t_{n} + (T-\Delta)/2) \\
& \; \; \: \; \; \;- H(t-t_{n} - (T-\Delta)/2))dt  \\
& = \int_{- \infty}^{\infty}N(t)\Pi\Big( \frac{t-t_{n}}{T-\Delta}  \Big) dt, \\
\end{split}
\end{equation}
where $H(t)$ is the Heaviside step function and we use $\Pi ((t-t_{n})/( T-\Delta))$ to denote a rectangular pulse of width $T-\Delta$ centered at $t_{n}$ with a normalized area of one. Noting $\Pi ((t-t_{n})/( T-\Delta)) \in \mathbb{R}$, we may re-express our average as a cross-correlation evaluated at time $t_{n}$,
\begin{equation}
N_{n} = [\Pi \Big( \frac{t-t_{n}}{ T-\Delta} \Big) \star N(t)] \Big|_{t=t_{n.}}
\end{equation}
The Fourier transform of $\Pi ((t-t_{n})/( T-\Delta))$ will incur a phase factor exp$(i \omega t_{n})$ resulting from the fact that the pulse is centered at $t_{n}$. We apply the convolution theorem again to express the frequency distribution of $N_{n}$ as
\begin{equation}
\label{Averaging_Effect_EVM_EQN}
\tilde{N}_{n}(f, S) = \tilde{\Pi}^{*}(f, S) \tilde{N}(f),
\end{equation}
While $\tilde{N}_{n}(f, S)$ will be complex-valued, EVM is calculated via the magnitude of the error vector lying in the complex plane. With this in mind, we take the absolute value of $\tilde{N}_{n}(f)$ to retrieve the magnitude of noise at a given frequency. In doing so, we use $|\tilde{\Pi}^{*}(s, S) | = | \tilde{\Pi} (f, S)|$, rendering the phase factor exp$(i \omega t_{n})$ inconsequential. As a result, $|\tilde{\Pi}(f,S)|$ is equivalent to the magnitude of the Fourier transform of $\Pi ( t/(T-\Delta))$, which we denote using $|\tilde{\Pi}_{0}(f,S)|$. Explicitly,
\begin{equation}
\tilde{\Pi}_{0}(f,S) = \frac{S}{\pi f(1-S\Delta)}sin \Big( \frac{\pi f(1-S\Delta)}{S} \Big).
\end{equation}
Using equations \ref{Multiplicative_EVM_EQN} and \ref{Averaging_Effect_EVM_EQN}, we conclude that non-atomic EVM for the $n^{th}$ symbol must scale with
\begin{equation} \label{Multiplicative_EVM_Contribution_EQN}
E_{n}(S) \propto \int_{- \infty}^{\infty} \tilde{B}(S) | \tilde{A}^{*}(f, S) \tilde{\Pi}_{0}(f, S) \tilde{N}(f) | df,
\end{equation}
where $\tilde{B}(S)$ refers to a spectral brickwall filter of width $2S$, centered on the peak of the central lobe of $\tilde{A}(f, S)$ (the positive frequency peak is used for heterodyne data). While the expression above is only for a single symbol, taking the magnitude of $\tilde{N}_{n}(f)$ has allowed us to ignore phase accumulation from $t_{n}$ as we move from one symbol to the next. In this case, the expression for any symbol in the transmission is identical to the expression above, meaning it effectively provides the total non-atomic EVM, $E(S)$.

We use equation \ref{Multiplicative_EVM_Contribution_EQN} to calculate  the relative quantity of non-atomic, multiplicative noise present in the EVM results. Since we take only the center point of the symbol duration, we use this result in the limit that $\Delta \rightarrow T = 1/S$, meaning the factor $\tilde{\Pi}_{0}(f, S)$ reduces to one. Since the ratio of atomic noise to non-atomic noise is unknown, we scale all the calculated values of $E(S)$ by a fixed constant which minimizes the difference in EVM between the datasets from each receiver, a choice motivated by the assumption that $E_{0}(S)$ is identical for both systems. Explicitly, figure (\ref{ComparisonFigure_Full}b) plots the result of equation \ref{Multiplicative_EVM_Contribution_EQN} for each measurement multiplied by this scalar value.

The fact that the removal of multiplicative pink noise contributions yields EVM results which are comparable between systems indicates that both the distribution (e.g., pink noise versus white noise) and type (additive versus multiplicative) directly influence which system may be more effective in demodulating a given signal. However, the fundamental limits of each system are still driven primarily by atomic noise, leaving open the possibility that a perfectly optimized DLRR could out-perform its conventional counterpart. In order to approach this idealized DLRR, $1/f$ laser intensity noise would need to be substantially reduced.

% Add additional discussion??? feedback systems which mitigate thermal effects in the diode which produces the probe laser light. Additionally, improvements such as the use of proportional-integral-derivative (PID) feedback could be used on the AOMs to ensure power drift in the laser beams is minimized. This would also reduce AT-splitting amplitude-based error, as coupling beam power drift would also be substantially reduced. 

\bibliography{bibliography}	
\end{document}